\RequirePackage{lineno} 

\documentclass[12pt,onecolumn,prc,showpacs,preprintnumber,amsmath,
floatfix,amssymb]{revtex4}
\usepackage{epsfig}
\usepackage{graphicx}
\usepackage{dcolumn}
\usepackage{bm}
\def\r{\mbox{{\bf  r}}}
\def\p{\mbox{\boldmath $p$}}
\def\q{\mbox{\boldmath $q$}}
\def\k{\mbox{\boldmath $k$}}
\def\t{\mbox{\boldmath $t$}}
\allowdisplaybreaks[4]

\begin{document}

\title{Neutrino neutral-current elastic scattering on ${}^{12}$C}
\author{A.~V.~Butkevich$^{1}$ and D.~Perevalov$^{2}$}
\affiliation{ 
$^1$Institute for Nuclear Research,
Russian Academy of Sciences,
60th October Anniversary Prosp. 7A,
Moscow 117312, Russia \\
$^2$Fermi National Accelerator Laboratory; Batavia, IL 60510
}
\date{\today}

\begin{abstract}

The neutral current elastic scattering of neutrinos on Carbon and $CH_2$ targets
 is computed using the relativistic distorted-wave impulse approximation with 
relativistic optical potential. Results for exclusive and inclusive neutrino 
reactions on ${}^{12}$C target are presented. We show that the nuclear effects 
on the shape of four-momentum transferred squared distribution 
$d\sigma/dQ^2_{QE}$ in neutrino neutral-current and charged-current 
quasi-elastic scattering are similar. We also calculate flux-averaged neutral 
current elastic differential cross section $d\sigma/dQ^2_{QE}$ for neutrino 
scattering from $CH_2$, as well as, the neutral-current to charged-current 
cross section ratio as functions of $Q^2_{QE}$. The value of axial mass 
$M_A$ is extracted from a fit of $d\sigma/dQ^2_{QE}$ cross section measured in 
MiniBooNE experiment. The extracted value of $M_A=1.28\pm 0.05$~GeV is 
consistent within errors with the MiniBooNE result. Additionally, for proton 
kinetic energies above the Cherenkov threshold, the strange quark contribution 
to the neutral current axial vector form factor at $Q^2_{QE}=0$, $\Delta s$, 
was extracted from a fit of MiniBoone data for 
$\nu p \to \nu p$ to $\nu N \to \nu N$ cross section ratio. This value is 
found to be $\Delta s=-0.11\pm 0.36$     
\end{abstract}
 \pacs{25.30.-c, 25.30.Bf, 25.30.Pt, 13.15.+g}

\maketitle

\section{Introduction}

Neutrino-nucleon neutral-current elastic (NCE) scattering provides an 
additional information about of structure of the hadronic weak neutral current 
(NC) and plays an important role in searching for the three active neutrino 
$\nu_{active}=\{\nu_e,\nu_{\mu},\nu_{\tau}\}$ conversion to a sterile neutrino 
$\nu_s$: a neutrino which has no coupling to neither W$^{\pm}$ nor Z$^{0}$ 
bosons.

The weak neutral current of the nucleon may be parametrized in terms of two 
vector and one axial-vector form factors. An additional induced pseudoscalar 
form factor is presented, but its contribution vanishes in the limit of a zero 
neutrino mass. In particular, the axial-vector form factor may be split into a 
non-strange and strange contributions. 
The latter one is proportional to the fraction 
of the nucleon spin carried by the strange quarks~\cite{Alberico1,Garvey1}. 
Thus the axial-vector form factor is crucial for understanding the role that
strange quarks play in determining the properties of nucleons.     

In order to investigate how the strange quarks contribute to the observed 
properties of the nucleon various reactions have been proposed: deep inelastic 
scattering of neutrino or polarized charged leptons on 
proton~\cite{Bazarko,Adams}, 
and parity-violating electron 
scattering~\cite{McKeown, Beck}. The strange vector form factors were 
measured in parity-violating electron scattering experiments~\cite{HAPPEX,
SAMPLE,A41,A42,G0}. A combined analysis of these experiments data points to 
small strangeness of the vector form factors~\cite{Liu}.     

Whereas parity-violating electron scattering is sensitive to the electric and 
magnetic strangeness, neutrino-induced reactions are sensitive to the strange 
quark contribution $\Delta s$ to the NC axial-vector form factor. A measurement
 of $\nu$($\bar{\nu}$)- proton NCE at Brookhaven National Laboratory (BNL E734)
~\cite{BNL} suggested a non-zero value of $\Delta s$. 
However, in Ref.~\cite{Garvey1} it has been shown that the BNL data cannot 
provide a decisive conclusion about the value of $\Delta s$ 
when taking into account uncertainties in the vector strange form-factors. 
Moreover, this result suffers strongly from experimental 
uncertainties due to difficulties in determination of the absolute neutrino 
flux. 

The measurement of the neutral-to-charged-current (CC) quasi-elastic 
cross section $R=NCE/CCQE$ 
in neutrino-nucleus scattering was proposed in Ref.~\cite{FINeSSE} to 
extract information on the strange spin of the proton 
because much of the systematic uncertainty is canceled by using the ratio. An 
important effort in this direction was the MiniBooNE experiment, that measured 
the flux-averaged NCE differential cross section $d\sigma/dQ^2$ as a function 
of four-momentum transferred squared $Q^2$ and ratio $R=NCE/CCQE$~\cite{MiniB}. 
The MINERvA experiment~\cite{MINERVA} aims at high precision measurements of 
neutrino scattering cross sections, and would be well-suited to examine the 
$Q^2$ evaluation of strangeness form factor in NCE scattering. 

Recently, the question about an additional sterile neutrino has drawn a 
considerable interest in the literature~\cite{Fogli,Donini,Dighe}. The 
short-baseline neutrino oscillation experiment, LSND, at the Los Alamos National 
Laboratory~\cite{LSND} reported evidence of 
$\bar{\nu}_{\mu}\rightarrow \bar{\nu}_e$ oscillation, but 
with a squared mass difference $\Delta m^2_{LSND}$ that 
is inconsistent within a three neutrino mass model with the two other values
extracted from solar, atmospheric, 
and reactor experiments, i.e. 
$\Delta m^2_{at} + \Delta m^2_{sol}\ne \Delta m^2_{LSND}$.

One of the most favorable scenarios that accommodates three independent 
$\Delta m^2$ values is an addition of a sterile neutrino. Because three 
active neutrinos couple to Z$^0$, the rate of neutrino NC events should be 
unaffected by the three flavor neutrino oscillation. Conversely, an existence 
of a sterile neutrino adds a possibility of a $\nu_{active}\rightarrow \nu_s$ 
transition that would create a deficit in the rate of NC events. 

However, the SNO experiment~\cite{SNO} made a neutral-current rate 
measurement and showed that the total flux of active neutrino from the Sun 
agree with expectation from the Standard Solar model. The Super-Kamiokande 
experiment excludes $\nu_{\mu}\rightarrow \nu_s$ and favors a pure 
$\nu_{\mu}\rightarrow \nu_{\tau}$ oscillation in its analysis of atmospheric 
neutrinos where an admixture of the two possibilities is 
allowed~\cite{SK1,SK2}. The MINOS collaboration reported~\cite{MINOS} the 
measurements of neutrino NC rates and spectra in an accelerator long baseline 
neutrino experiment. The rates at the near and far detectors are consistent 
with expectations from decay kinematics and geometry, providing new support 
for the interpretation of muon neutrino disappearance as oscillations among 
the three active neutrinos. 
So, an additional interest in the neutrino-nucleus 
NCE scattering cross section is that this process plays a key role in a search 
for the parameter space available for $\nu_{active}\rightarrow \nu_s$ 
oscillations.  

It has been shown in Refs.~\cite{Denis,MiniB,Benhar} that 
in order to measure the strange quark contribution
to the nucleon spin using a neutrino-nucleon NCE cross section
it is necessary to distinguish $\nu p\to\nu p$ from $\nu n\to\nu n$ 
interactions.Otherwise, the total NCE cross-section on both proton and neutron 
($\nu N\to\nu N$)has a negligible dependence on the nucleon spin's strangeness.
A detailed analysis of the NCE scattering cross section's sensitivity 
to the strange content of nucleon neutral current was carried out
in a relativistic plane-wave impulse approximation in Ref.~\cite{Ventel}.

Analysis of nuclear structure effects on the determination of the strange quark 
contribution in neutrino-nucleus NCE scattering were performed in 
Refs.\cite{Alberico2,Barbado} where the relativistic Fermi Gas model 
(RFGM) and relativistic shell model including final state interaction (FSI) of 
outgoing nucleon were used. The effects of FSI were also studied in 
Ref.~\cite{Bleve} within the RFGM and in Refs.~\cite{Garvey1,Botrugno} in the 
framework of the Random Phase Approximation theory. 
The effects of FSI on the ratio of proton-to-neutron 
cross section in NCE scattering were discussed in Refs.\cite{Alberico2,
Horowitz,Garvey2}.  

The effects of FSI on NCE scattering cross 
section were studied in Refs.~\cite{Meucci1,Meucci2,Kim} within the framework 
of a relativistic distorted-wave impulse approximation (RDWIA) with a 
relativistic optical potential. In Refs.~\cite{Meucci1,Meucci2} 
important FSI effects arise from the use of optical potential within a 
relativistic Green's function approach. An analys of the sensitivity of NCE 
scattering cross section to the strangeness contribution was presented also in 
Refs.\cite{Martinez,Jachowicz} within the RDWIA and relativistic 
multiple-scattering Glauber approximation.

In this paper we present the RDWIA calculation of the neutrino-nucleon NCE 
scattering cross section on Carbon and $CH_2$. In this approach that was 
successfully applied in 
Refs.\cite{BAV1,BAV2,BAV3,BAV4} to CC quasi-elastic scattering we 
calculated the flux-averaged $d\sigma/dQ^2$ cross section and ratio 
$R(NCE/CCQE)$ and compare the results with the MiniBooNE data~\cite{MiniB}.
Additionally, the ratio of the predicted event rates in the MiniBooNE 
high energy $\nu p \to nu p$ and $\nu N \to nu N$ event samples was calculated.
This ratio is sensitive to the strange quark contribution to the nucleon spin, $\Delta s$.
Using the MiniBooNE data for this distribution we 
performed a measurement of $\Delta s$ and compared it to the MiniBooNE result reported in Ref.~\cite{MiniB}.

The outline of this article is as follows: In Sec. II we present briefly the
formalism for the NCE scattering process and the RDWIA approach. The results 
are presented and discussed in Sec. III. Our conclusions are summarized in 
Sec. IV.

\section{The formalism and model for the neutral-current elastic scattering}

In this section we consider the formalism for description of NCE exclusive
\begin{equation}\label{Eq.1}
\nu(k_i) + A(p_A)  \rightarrow \nu(k_f) + N(p_x) + B(p_B),      
\end{equation}
and inclusive
\begin{equation}\label{Eq.2}
\nu(k_i) + A(p_A)  \rightarrow \nu(k_f) + X                      
\end{equation}
scattering off nuclei in the one-$Z^0$-boson exchange approximation. Here 
$k_i=(\varepsilon_i,\k_i)$ 
and $k_f=(\varepsilon_f,\k_f)$ are the initial and final lepton 
momenta, $p_A=(\varepsilon_A,\p_A)$, and $p_B=(\varepsilon_B,\p_B)$ are 
the initial and final target momenta, $p_x=(\varepsilon_x,\p_x)$ is the 
ejectile nucleon momentum, $q=(\omega,\q)$ is the momentum transfer carried by 
the virtual $Z^0$-boson, and $Q^2=-q^2=\q^2-\omega^2$ is the $Z^0$-boson 
virtuality. As the basic outline follows closely the CC formalism developed in 
Ref.~\cite{BAV1}, we present a brief review that focuses on those 
modifications that arise from the weak neutral current.

\subsection{Neutrino-nucleus NCE scattering cross sections} 

In the laboratory frame, the differential cross section for the exclusive
(anti-)neutrino NCE scattering, in which only a single 
discrete state or narrow resonance of the target is excited, can be written as
\begin{equation}
\label{Eq.3}
\frac{d^5\sigma^{(nc)}}{d\varepsilon_f d\Omega_f d\Omega_x} = R
\frac{\vert\p_x\vert{\varepsilon}_x}{(2\pi)^5}\frac{\vert\k_f\vert}
{\varepsilon_i} \frac{G^2}{2} L_{\mu \nu}^{(nc)}W^{\mu \nu (nc)},
\end{equation}
 where $\Omega_f$ is the solid angle for the lepton momentum, $\Omega_x$ is the
 solid angle for the ejectile nucleon momentum, 
$R$ is the recoil factor, $G \simeq 1.16639 \times 10^{-11}$~MeV$^{-2}$ is
the Fermi constant, $L^{(nc)}_{\mu \nu}$ and $W^{(nc)}_{\mu \nu}$ are NC lepton and 
nuclear tensors, respectively. 

The energy $\varepsilon_x$ is the solution to the equation
\begin{equation}\label{Eq.4}
\varepsilon_x+\varepsilon_B-m_A-\omega=0,                                 
\end{equation}
where $\varepsilon_B=\sqrt{m^2_B+\p^2_B}$, $~\p_B=\q-\p_x$, $~\p_x=
\sqrt{\varepsilon^2_x-m^2}$, and $m_A$, $m_B$, and $m$ are masses of the 
target, recoil nucleus and nucleon, respectively. 
The missing momentum $p_m$ and missing energy $\varepsilon_m$ are defined by 
\begin{subequations}
\begin{align}
\label{Eq.5}
\p_m & = \p_x-\q
\\
\label{eps_m}
\varepsilon_m & = m + m_B - m_A                                           
\end{align}
\end{subequations}
From Eq.\eqref{Eq.4} the total energy of the ejected nucleon is given by
\begin{equation}\label{Eq.6}
\varepsilon_x=\omega + m_A - \varepsilon_B \approx \omega + m - 
(\varepsilon_m-p^2_m/2m_B)                                 
\end{equation}
and the nucleon kinetic energy can be written as
\begin{equation}\label{Eq.7}
T_N=\omega - (\varepsilon_m-p^2_m/2m_B) \approx \omega - \varepsilon_m,     
\end{equation}
if one neglects the recoil nucleon energy $p^2_m/2m_B$. As the outgoing neutrino 
is undetected the differential cross section Eq.\eqref{Eq.3} can be rewritten 
in ``no-recoil'' approximation as follows
\begin{equation}\label{Eq.8}
\frac{d^5\sigma^{(nc)}}{dT_N d\Omega_fd\Omega_x} \approx \frac{d^5\sigma^{(nc)}} 
{d\varepsilon_f d\Omega_f d\Omega_x}
\end{equation}

The leptonic tensor $L^{(nc)}_{\mu \nu}$ is separated into symmetric and 
antisymmetric components that are given as in Ref.~\cite{BAV1}. Note, that 
the weak lepton NC is conserved for massless neutrino and $q^{\mu}
L^{(nc)}_{\mu\nu}=L^{(nc)}_{\mu\nu}q^{\nu}=0$. All the nuclear structure information 
and FSI effects are contained in the weak NC nuclear 
tensor $W^{(nc)}_{\mu \nu}$, which is given by the bilinear product of the 
transition matrix elements of the nuclear NC operator $J^{(nc)}_{\mu}$ between 
the initial nucleus state $|A\rangle$ and the final state $|B_f\rangle$ as 
\begin{eqnarray}
\label{Eq.9}
W^{(nc)}_{\mu \nu } &=& \sum_f \langle B_f,p_x\vert                           
J^{(nc)}_{\mu}\vert A\rangle \langle A\vert
J^{(nc) \dagger}_{\nu}\vert B_f,p_x\rangle,              
\label{W}
\end{eqnarray}
where the sum is taken over undetected states. 
This tensor is an extremely
complicated object as, in principle, 
the exact form for many body wave functions and 
operators must be used. A general model-independent covariant form of 
$W^{(nc)}_{\mu\nu}$ and the result of its contraction with the leptonic tensor were 
obtained in Ref.~\cite{Ventel}. There it was shown that the contraction 
$L_{\mu \nu}^{(nc)}W^{\mu\nu (nc)}$ and therefore the differential cross 
section in Eq.~\eqref{Eq.3} is completely determined by a set of eight structure 
functions.  

General expressions for the cross sections of the exclusive and inclusive 
CCQE neutrino scattering off nucleus are given in Ref.~\cite{BAV1} in terms of 
weak response functions. 
In order to apply these expressions for calculation of 
neutrino-nucleus NCE scattering cross sections it is necessary to replace 
$G^2\cos^2\theta_C \to G^2$, express the response functions as suitable 
combinations of the hadron tensor components $W^{(nc)}_{\mu\nu}$, and 
calculate the coefficient $v_i$ for massless neutrino. 
The single differential cross section as a function of
the outgoing nucleon's kinetic energy $T_N$ can be 
obtained after performing integration of the cross section in Eq.\eqref{Eq.8} 
over solid angles of the outgoing neutrino and nucleon. 

\subsection{Model}

We describe the neutrino-nucleon NCE scattering in the impulse approximation,
assuming that the incoming neutrino interacts with only one nucleon, which is 
subsequently emitted, while the remaining ($A$-1) nucleons in the target are 
spectators. When the nuclear current is written as the sum of single-nucleon 
currents, the nuclear matrix element in Eq.\eqref{Eq.9} takes the form
\begin{eqnarray}\label{Eq.10}
\langle p,B\vert J^{\mu (nc)}\vert A\rangle &=& \int d^3r~ \exp(i\t\cdot\r)
\overline{\Psi}^{(-)}(\p,\r)
\Gamma^{\mu (nc)}\Phi(\r),                                          
\end{eqnarray}
where $\Gamma^{\mu (nc)}$ is the NC vertex function, $\t=\varepsilon_B\q/W$ is the
recoil-corrected momentum transfer, $W=\sqrt{(m_A+\omega)^2-\q^2}$ is the
invariant mass, $\Phi$ and $\Psi^{(-)}$ are the relativistic bound-state and
outgoing wave functions.

The single-nucleon charged current has a $V{-}A$ structure $J^{(nc)\mu} = 
J^{\mu (nc)}_V + J^{\mu (nc)}_A$. For a free-nucleon vertex function, 
$\Gamma^{\mu (nc)} = \Gamma^{\mu (nc)}_V + \Gamma^{\mu (nc)}_A$, we use the vector 
current vertex function 
\begin{equation}\label{Eq.11}
\Gamma^{\mu (nc)}_V = F^{(nc)}_V(Q^2)\gamma^{\mu} + 
{i}\sigma^{\mu \nu}q_{\nu}F^{(nc)}_M(Q^2)/2m,                      
\end{equation}
and the axial current vertex function
\begin{eqnarray}\label{Eq.12}
\Gamma^{\mu (nc)}_A = F^{(nc)}_A(Q^2)\gamma^{\mu}\gamma_5 +         
F^{(nc)}_P(Q^2)q^{\mu}\gamma_5.
\end{eqnarray}

The vector form factors $F^{(nc)}_i$ ($i=V,M$) are related to the corresponding 
electromagnetic ones for proton $F^p_i$ and neutron $F^n_i$, plus a possible 
isoscalar strange-quark contribution $F^s_i$, i.e.~\cite{Alberico1} 
\begin{subequations}
\begin{align}
\label{Eq.13}
F^{(nc)}_V & = \tau_3(0.5-\sin^2\theta_W)(F^p_1 - F^n_1)           
- \sin^2\theta_W(F^p_1 + F^n_1) - F^s_V/2
\\
F^{(nc)}_M & = \tau_3(0.5-\sin^2\theta_W)(F^p_2 - F^n_2) 
 - \sin^2\theta_W(F^p_2 + F^n_2) - F^s_M/2,
\end{align}
\end{subequations}
where $\tau_3=+(-1)$ for proton (neutron) knockout and $\theta_W$ is the 
Weinberg angle ($\sin^2\theta_W\approx 0.2313$). The axial $F_A^{(nc)}$ form 
factor is expressed as 
\begin{eqnarray}\label{Eq.14}
\Gamma^{\mu (nc)}_A = (\tau_3 F_A - F^s_A)/2,                        
\end{eqnarray}
where $F^s_A$ describes possible strange-quark contributions. In this work we 
neglect the strangeness contributions, i.e., it is supposed that 
$F^s_V=F^s_M=F^s_A=0$. For the nucleon form factors $F^{p(n)}_i$ the 
approximation of Ref.~\cite{MMD} is used. Because the bound nucleons are the 
off-shell we employ the de Forest prescription~\cite{deFor} and Coulomb gauge 
for off-shell vector current vertex $\Gamma^{\mu}_V$. The vector-axial 
form factor is parametrized as a dipole with the axial nucleon mass $M_A$, 
which controls the $Q^2$ dependence of $F_A$.

The independent particle shell model (IPSM) is assumed in the calculations of 
the nuclear structure, taking into account the short-range nucleon-nucleon 
($NN$) correlation in the ground state.
According to the experimental data~\cite{Dutta,Kelly1} the occupancy of the 
IPSM orbitals of ${}^{12}$C equals on average 89\%. We assume that the missing 
strength (11\%) can be attributed to the $NN$ correlations, leading to the 
appearance of the high-momentum and high-energy component in the nucleon 
distribution in the target. To estimate this effect in the inclusive cross 
sections, we consider a phenomenological model that incorporates both the 
single-particle nature of the nucleon spectrum at low energy (IPSM orbitals) 
and the high-energy and high-momentum components due to $NN$ correlations.

For ${}^{12}$C we use the same relativistic wave functions of the 
bound nucleon states $\Phi$ as in Refs.\cite{BAV3,BAV4}. The wave functions 
were obtained from Ref.~\cite{TIMORA} as the self-consistent solutions of the 
relativistic Hartree-Bogoliubov equations, derived within a relativistic mean 
field approach. The normalization factors $S(\alpha)$ relative to the 
full occupancy of the IPSM orbitals of ${}^{12}$C~\cite{Dutta,Kelly1} are: 
$S(1p_{3/2})$=84\%, $S(1s_{1/2})$=100\% with an average factor of about 89\%. 

In order to take into account FSI effects in the RDWIA, the distorted-wave 
function, $\Psi$, is evaluated as a solution of the 
Dirac equation containing a phenomenological relativistic optical potential. 
The channel coupling in the FSI~\cite{Kelly2} of the $N+B$ system is taken into 
account. The relativistic optical potential consists of a real part which 
describes rescattering of the ejected nucleon and an imaginary part that 
accounts for its absorption into unobserved channels.
We use the LEA program~\cite{LEA} for the numerical calculation of 
the distorted wave functions with the EDAD1 parametrization~\cite{Cooper} of 
the relativistic optical potential for Carbon. This code, initially designed 
for computing exclusive electron-nucleus scattering, was 
successfully tested against A$(e,e'p)$ data~\cite{Fissum,Dutta}, and adopted 
for neutrino reactions~\cite{BAV1}.  

A complex optical potential with a nonzero imaginary part generally produces 
an absorption of the flux. For the exclusive $A(l,l'N)$ channel this reflects 
the coupling between different open reaction channels. However, for the 
inclusive reaction, the total flux must be conserved. 
In Refs.~\cite{Meucci3,Meucci4}, it was shown that the inclusive CCQE neutrino 
cross section of the exclusive channel $A(l,l'N)$ calculated with only the real 
part of the optical potential is almost identical when calculated via the 
Green's function approach~\cite{Meucci3}, in which the FSI effects on 
inclusive reaction $A(l,l'X)$ are treated by means of a complex potential, and 
the total flux is conserved. We calculate the inclusive $d\sigma/dQ^2$ 
sections with the EDAD1 relativistic optical potential in which only the real 
part is included. The inclusive cross sections with FSI effects in the 
presence of short-range $NN$ correlations were calculated using the method 
proposed in Ref.~\cite{BAV1}. 

\section{Results and discussion}

\subsection{Neutral Current Elastic differential cross section}

The exclusive reaction $\nu +A \to \nu + p +B$ reaction is a good signal sample
 of (anti)neutrino NCE scattering off nuclei. The measurement of the range of 
the scattered proton, its angle with respect to the direction of the incident 
neutrino $\theta_p$, and its rate of energy loss allow to identify the particle
 as a proton and determine proton kinetic energy $T_p$. In impulse 
approximation, assuming the target nucleon to be at rest inside nucleus, these 
measured quantities $(T_p, \cos\theta_p)$ determine the neutrino energy through
 the kinematic relation 
\begin{eqnarray}
\label{Eq15}
\varepsilon_{\nu}=\frac{m_p}{\cos\theta_p(1+2m_p/T_p)^{1/2}-1},
\end{eqnarray}
where $m_p$ is the proton mass.
\begin{figure*}
  \begin{center}
    \includegraphics[height=16cm,width=16cm]{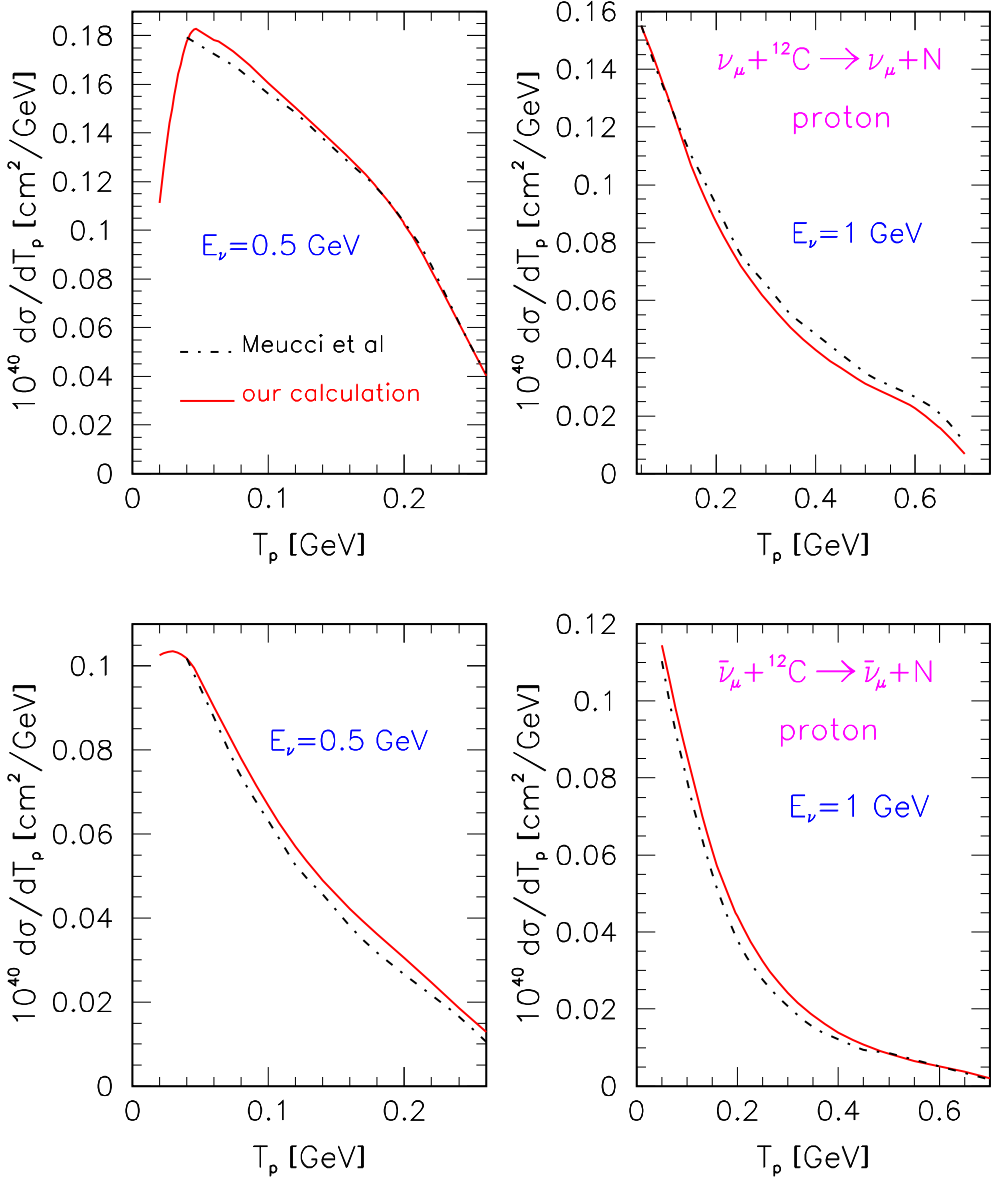}
  \end{center}
  \caption{\label{Fig.1}(Color online) Differential cross sections $\sigma_p$ of 
neutrino (upper panel) and antineutrino (lower panel) NCE scattering as a 
function of the outgoing proton kinetic energy for two the values of incoming 
(anti)neutrino energy: $\varepsilon_{\nu}=500$ and $1000$~MeV,   
calculated in the RDWIA. The solid lines represent the results obtained in 
this work and the dot-dashed lines are the results of Ref.~\cite{Meucci1}.
}
\end{figure*}
In neutrino oscillation experiments with two detectors the spectra of the 
protons as functions of neutrino energies, measured at near and far detectors 
can be used to search for $\nu_{active}\to\nu_s$ transition that would create a 
deficit in the rate of one proton events at far detector. Note, that precise
 measurement of the (anti)neutrino NCE scattering off neutron appear 
problematic due to the difficulties associated with neutron detection.
\begin{figure*}
  \begin{center}
    \includegraphics[height=16cm,width=16cm]{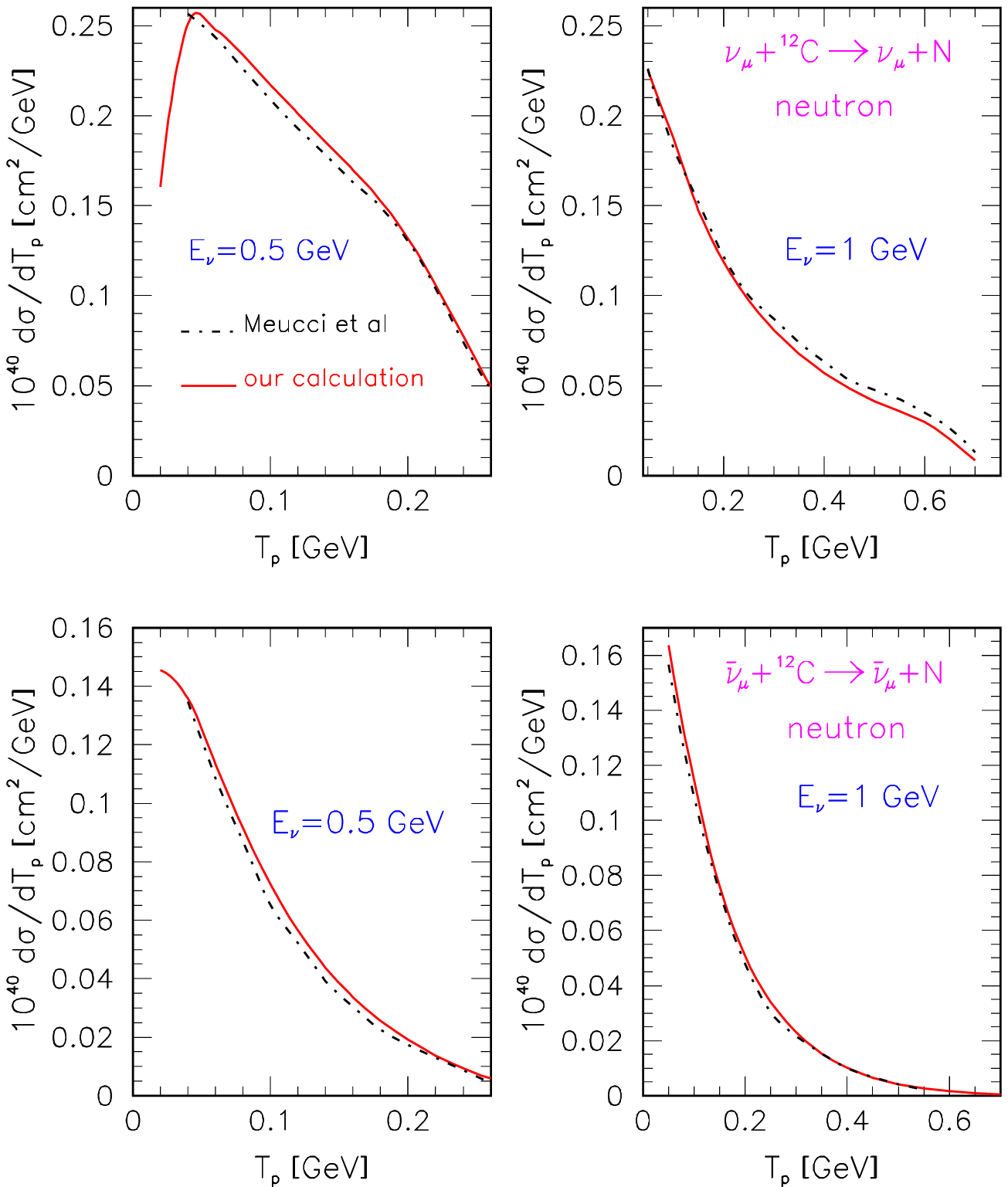}
  \end{center}
  \caption{\label{Fig.2}(Color online) The same as Fig.\ref{Fig.1}, but for 
neutron knockout ($\sigma_n$). 
} 
\end{figure*}
\begin{figure*}
  \begin{center}
    \includegraphics[height=16cm,width=16cm]{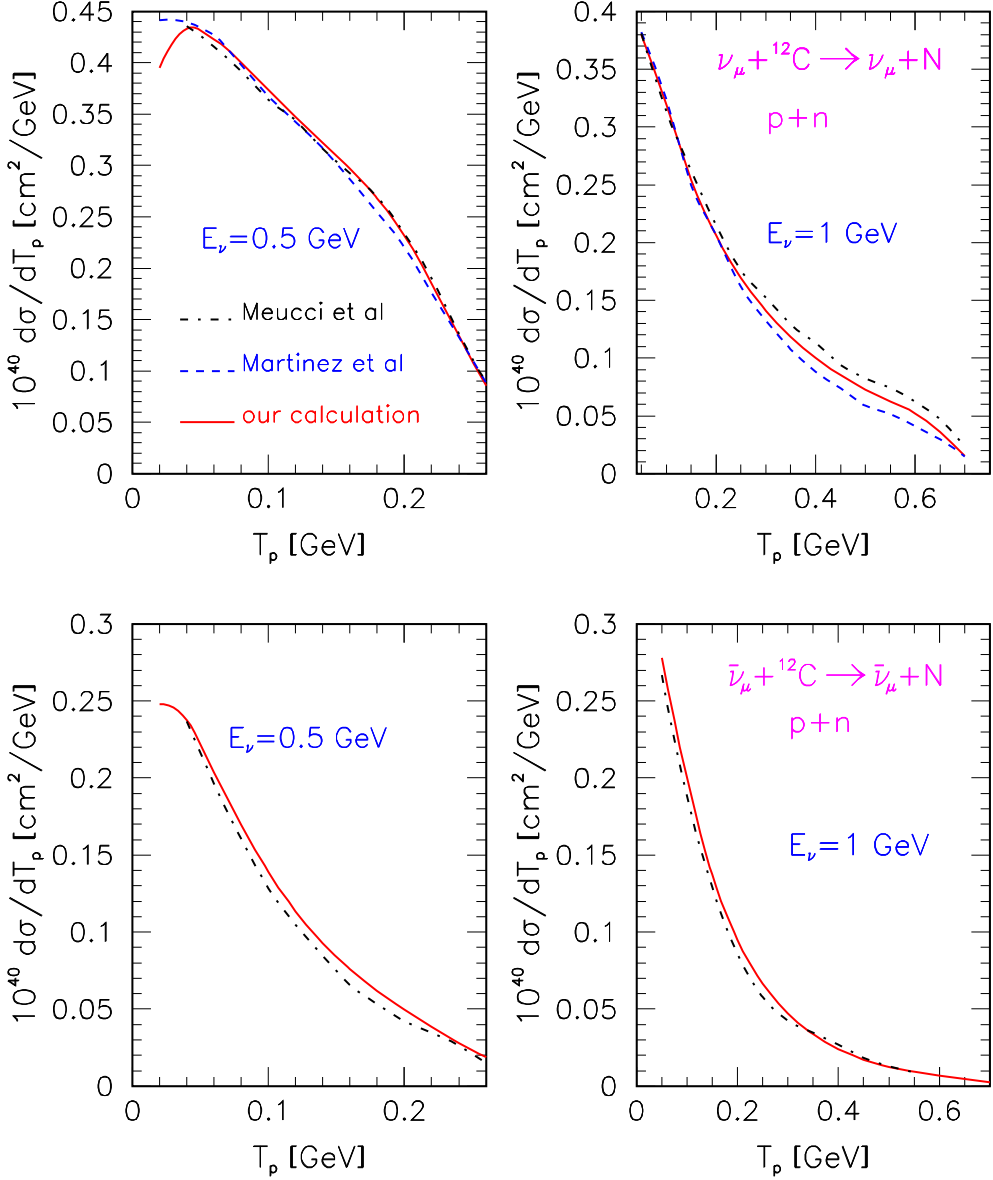}
  \end{center}
  \caption{\label{Fig.3}(Color online) The same as Fig.\ref{Fig.1}, but for 
proton or neutron knockout ($\sigma_p + \sigma_n$). The dashed lines show the 
RDWIA results of Ref.~\cite{Martinez}. 
} 
\end{figure*}
\begin{figure*}
  \begin{center}
    \includegraphics[height=16cm,width=16cm]{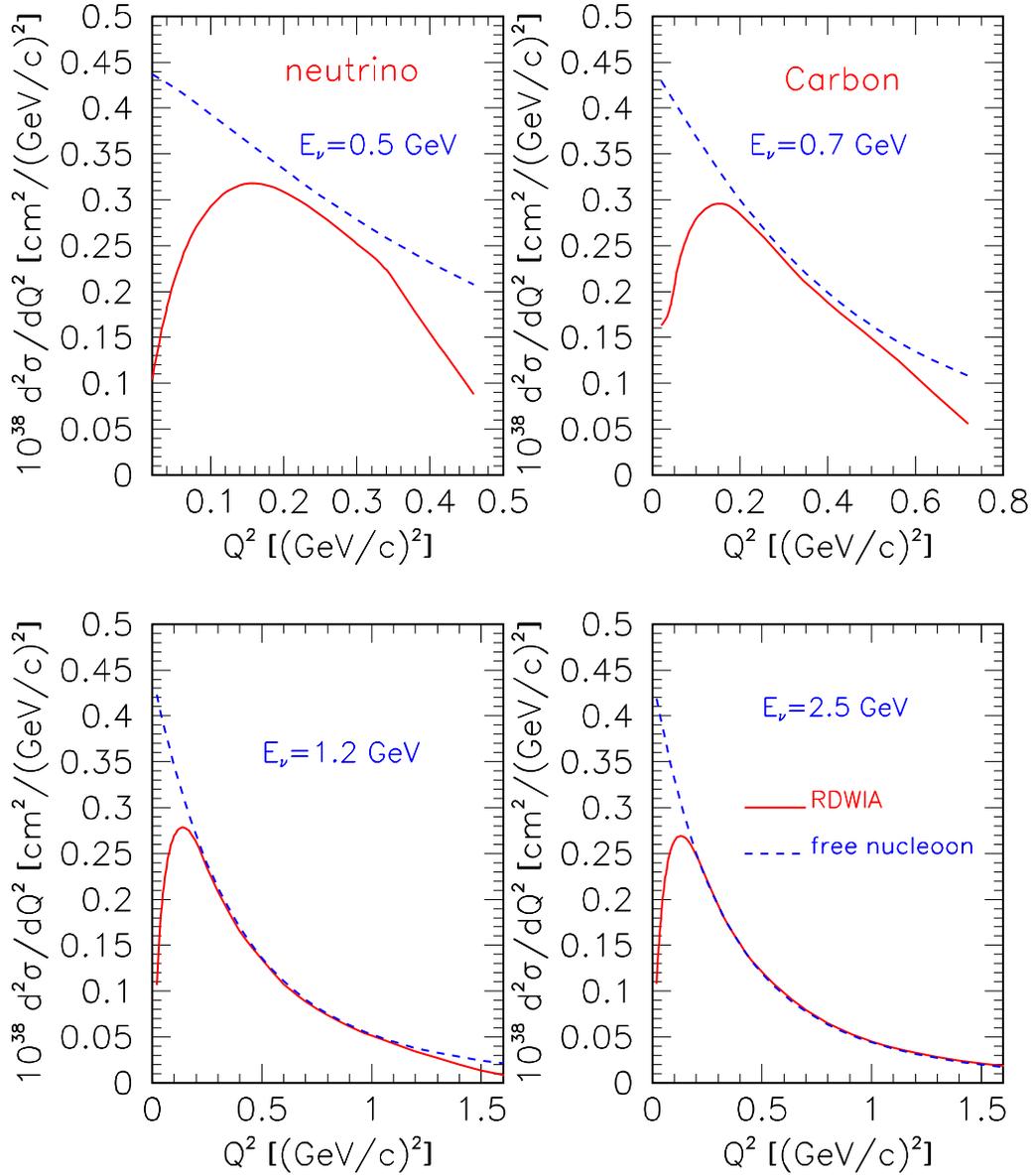}
  \end{center}
  \caption{\label{Fig.4}(Color online) Inclusive NCE cross sections vs. the 
four-momentum transfer $Q^2$ for neutrino scattering off ${}^{12}$C 
(solid line) and free nucleon (dashed line) and for the four values of 
incoming neutrino energy: $\varepsilon=0.5, 0.7, 1.2$ and $2.5$~GeV.
}
\end{figure*}
\begin{figure*}
  \begin{center}
    \includegraphics[height=16cm,width=16cm]{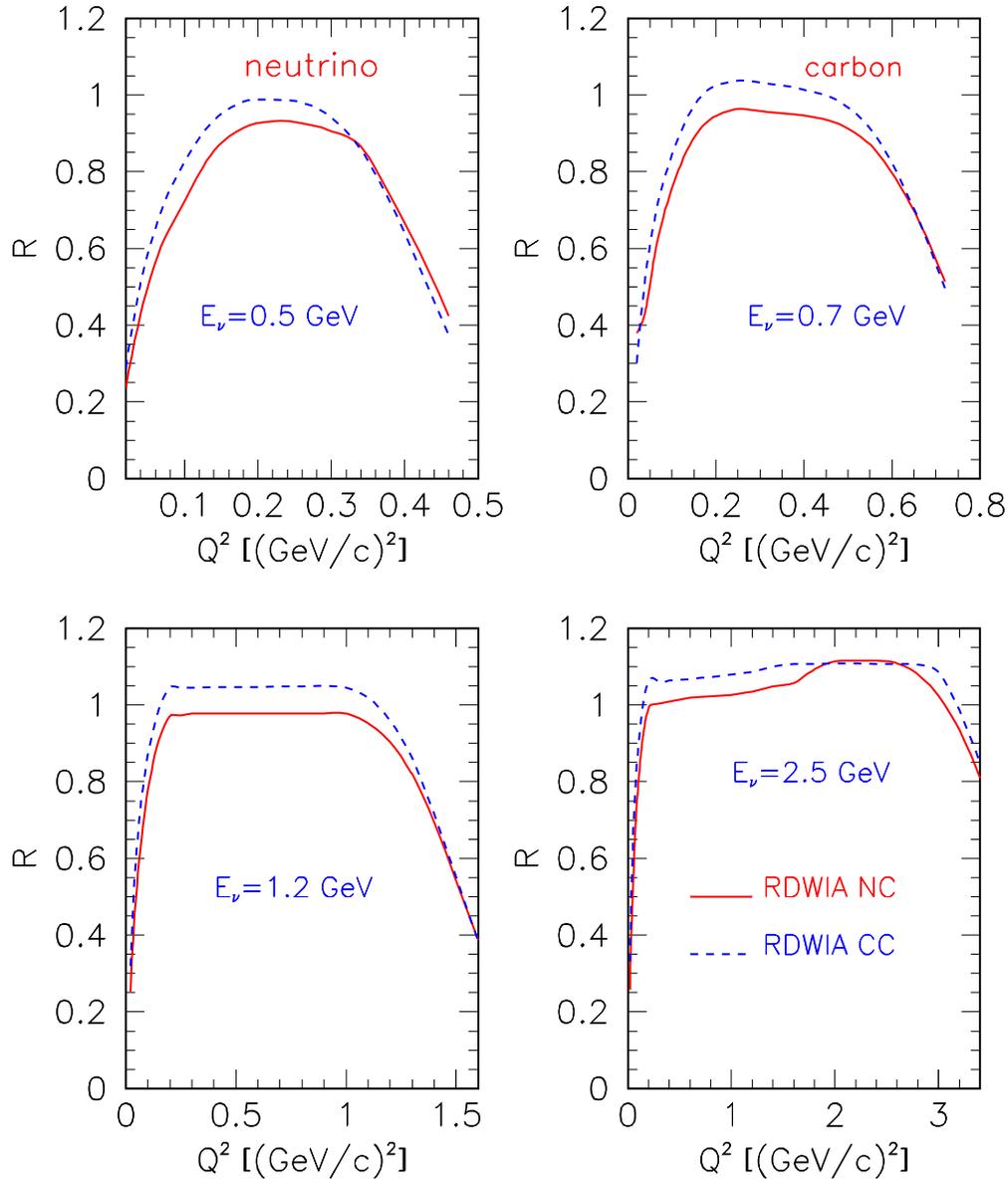}
  \end{center}
  \caption{\label{Fig.5}(Color online) Ratio $R(\varepsilon_{\nu},Q^2)$ vs. the 
four-momentum transfer $Q^2$ for neutrino scattering off ${}^{12}$C and for the 
four values of incoming neutrino energy: $\varepsilon=0.5, 0.7, 1.2$ and $2.5$ 
GeV. As shown in the key, the ratios were calculated for neutrino NCE and CCQE
 scattering.
}
\end{figure*}
\begin{figure*}
  \begin{center}
    \includegraphics[height=16cm,width=16cm]{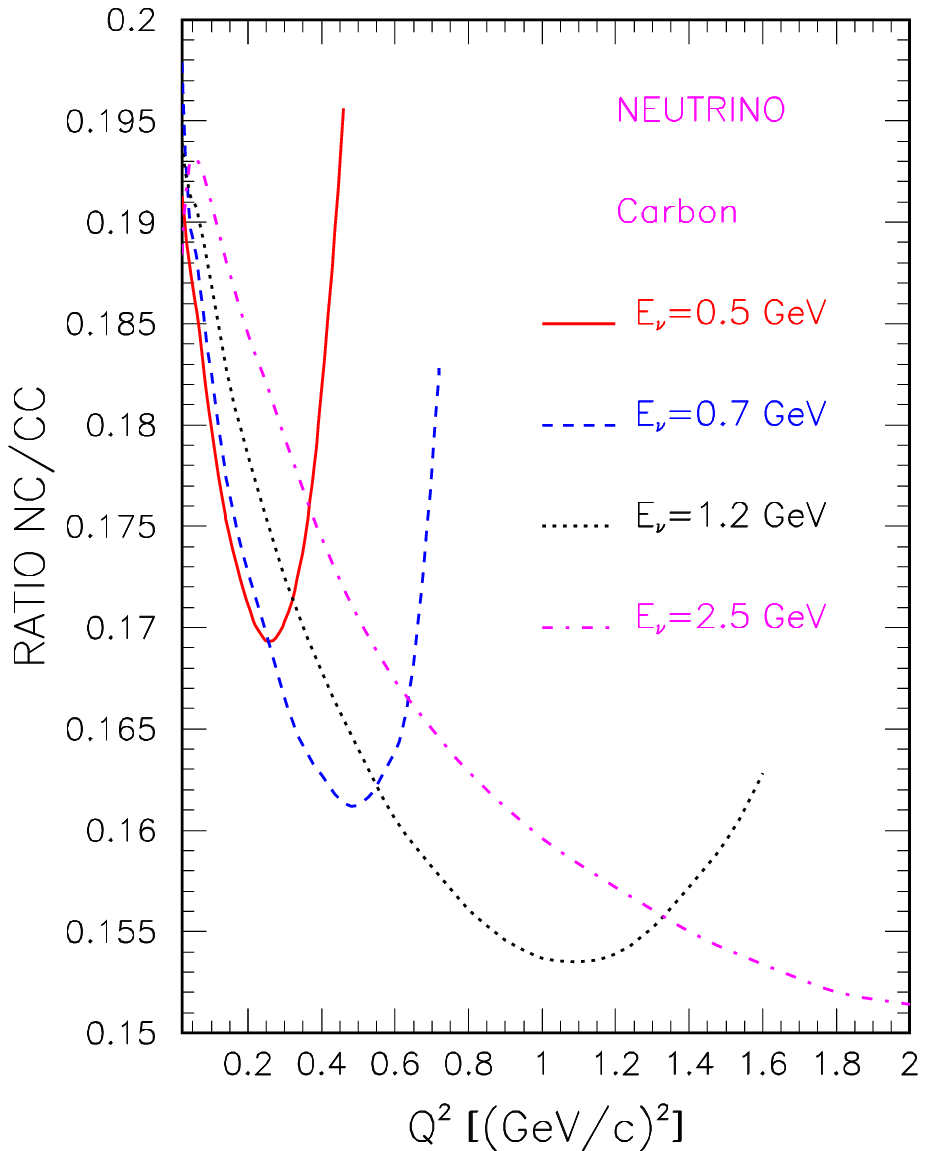}
  \end{center}
  \caption{\label{Fig.6}(Color online) Ratio of neutral-to-charged-current 
cross sections $R=NC/CC$ vs. the four-momentum transfer $Q^2$ for neutrino 
scattering off ${}^{12}$C and for the four values of incoming neutrino energy: 
$\varepsilon=0.5, 0.7, 1.2$ and $2.5$~GeV. 
}
\end{figure*}
In Figs.\ref{Fig.1},~\ref{Fig.2}, and \ref{Fig.3} the (anti)neutrino NCE 
exclusive cross sections $\sigma_P=d\sigma_p/dT_p$ (per bound proton), 
$\sigma_n=d\sigma_n/dT_n$ (per bound neutron), and sum $\sigma_p + \sigma_n$ are 
displayed as functions of the emitted nucleon kinetic energy for proton, 
neutron, and proton or neutron knockout, respectively. The calculations 
correspond to Carbon target and incoming energies of 500~MeV and 1000~MeV. The
 upper (lower) panels show the cross sections for neutrino (antineutrino) NCE 
scattering in comparison with results obtained in Refs.\cite{Martinez, 
Meucci1}. These cross sections were also calculated in the RDWIA approach with 
dipole approximation of the nucleon form factors, EDAD1 parametrization of 
relativistic optical potential and neglecting the $NN$ correlations in the 
ground state of Carbon. We observe that our calculations and those 
performed in the framework of RDWIA formalism are in a good agreement. 
For neutrino (antineutrio) 
the $\sigma_p/\sigma_n$ NCE cross section ratio increases almost linearly
with nucleon energy: from $\approx 0.7~(\approx 0.7)$ for 
$T_N\approx 20-50$~MeV up to $\approx 0.82~(\approx 2)$ for $T_N=700$~MeV. 

To study the nuclear effects on the $Q^2$ distribution, we calculated (with 
$M_A=1.032$~GeV and $F^s_A=0$) the inclusive cross sections 
$(d\sigma/dQ^2)_{nuc}$ (per bound nucleon) of the neutrino NCE scattering on 
Carbon. The results for neutrino energies $\varepsilon_{\nu}=0.5, 0.7, 1.2$ and 
$2.5$~GeV are shown in Fig.\ref{Fig.4} in comparison with cross section for 
neutrino NCE scattering on a free nucleon 
$(d\sigma/dQ^2)_{free}=0.5[(d\sigma/dQ^2)_p + (d\sigma/dQ^2)_n]$, 
where $(d\sigma/dQ^2)_p$ and $(d\sigma/dQ^2)_n$ are the cross sections for 
neutrino NCE scattering on free proton and neutron, respectively. 

Nuclear effects on the shape of the $Q^2$ distribution, 
i.e., ratios $R(\varepsilon_{\nu},Q^2)=(d\sigma/dQ^2)_{nuc}/(d\sigma/dQ^2)_{free}$ 
are presented in Fig.\ref{Fig.5}. The results obtained for neutrino energies 
$\varepsilon_{\nu}=0.5, 0.7, 1.2$ and $2.5$~GeV are compared with those 
calculated for neutrino CCQE scattering in Ref.~\cite{BAV3}. We observe that 
nuclear effects in neutrino NCE and CCQE scattering, in general, are similar. 
The nuclear effects are seen at low $Q^2$; the tail of the momentum 
distribution at high $Q^2$, an overall suppression, and slight change in slope 
in the middle region at $\varepsilon_{\nu} \ge 1$~GeV is also observed. 
The range of $Q^2$ where $R\approx 1$ (i.e., nuclear effects are small 
and therefore cannot affect the measurement of the effective $M_A$) 
increases with the incoming neutrino energy.    

The measurement of the neutral-to-charged-current cross sections ratio in the 
neutrino-nucleus scattering was proposed in Ref.~\cite{FINeSSE} to extract 
a possible strange-quark contribution. Our RDWIA results for  
$R=NC/CC=(d\sigma/dQ^2)^{NC}/(d\sigma/dQ^2)^{CC}$ ratio, obtained with 
$M_A=1.032$~GeV and $F^s_A=0$ are presented in Fig.\ref{Fig.6} as functions of 
$Q^2$ for neutrino energies $\varepsilon=0.5, 0.6, 1.2$ and $2.5$~GeV. The 
inclusive CCQE cross sections $(d\sigma/dQ^2)^{CC}$ were in Ref.~\cite{BAV3}. 
The $NC/CC$ ratio decreases as $Q^2$ increases from $\sim 1.9$ at 
$Q^2\approx 0.1$ (GeV/c)$^2$ and reaches the minimum at large value of $Q^2$. 
The fact that the CCQE cross section goes to zero more rapidly than the 
corresponding NCE one (due to the muon mass) causes the enhancement of the 
ratio at large value of $Q^2$ close to the upper border of the allowed 
kinematic range of $Q^2$. The results obtained in Refs.\cite{Meucci1,Martinez} 
show similar features.
 
\subsection{MiniBooNE flux-averaged differential cross section: comparison 
with data}

The MiniBooNE collaboration reported~\cite{MiniB} high-statistic 
measurement of the flux-averaged NCE differential cross section for neutrino 
scattering on $CH_2$ as a function of $Q^2_{QE}$. 
\begin{figure*}
  \begin{center}
    \includegraphics[height=16cm,width=16cm]{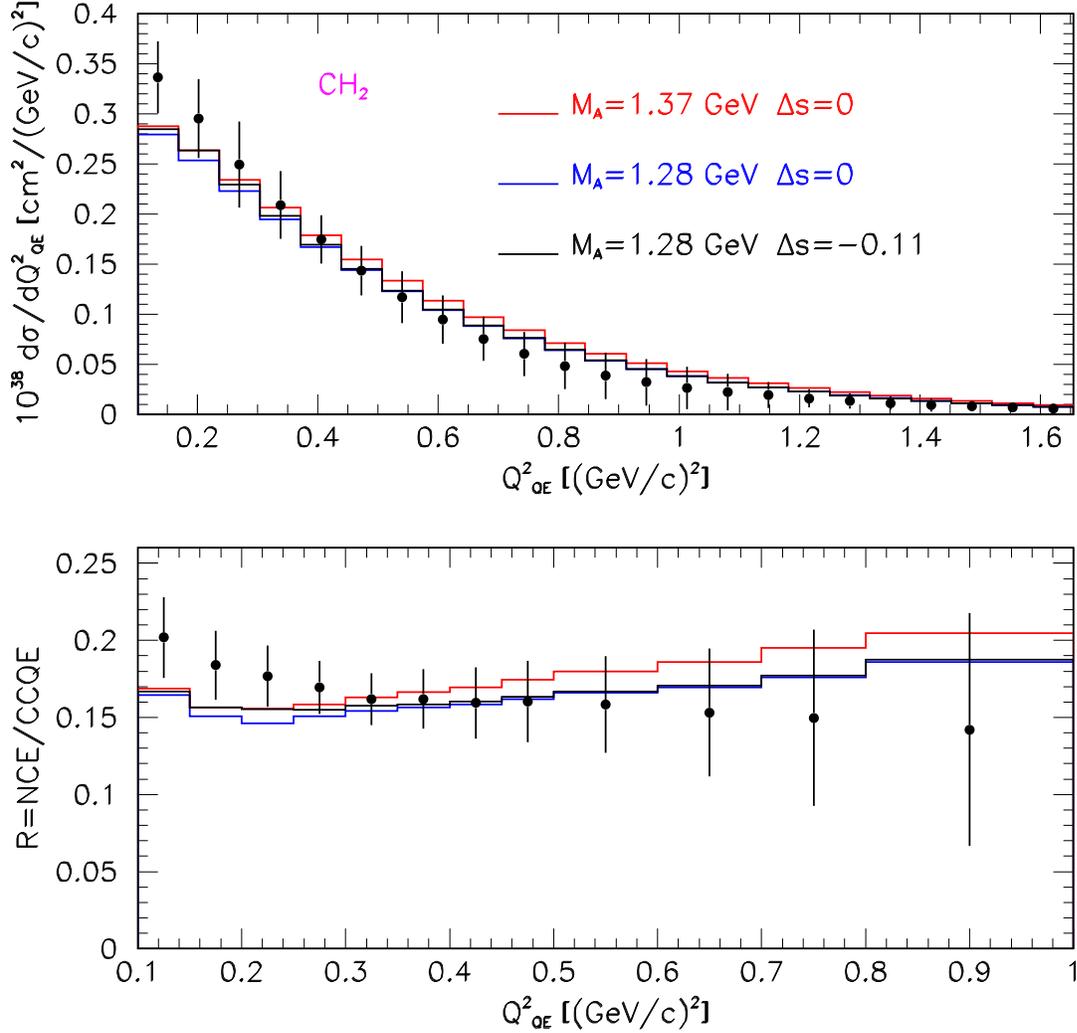}
  \end{center}
  \caption{\label{Fig.7}(Color online) Flux-averaged $d\sigma/dQ^2_{QE}$ 
cross section per nucleon (upper panel) for neutrino scattering on $CH_2$ 
and $NCE/CCQE$ cross section ratio (lower panel) as a function of $Q^2_{QE}$. 
The NCE cross section and CCQE/NCE ratio calculated with values of $M_A=1.28$~GeV 
(blue line), 1.37~GeV (red line), both with the value $\Delta s = 0.0$.
Also shown is the strange quark effect on the NCE cross-section and the ratio with a
value of $\Delta s=-0.11$ and $M_A=1.28$~GeV (black line).
The MiniBooNE data are shown as points.  
}
\end{figure*}
In this experiment the sum of kinetic energies of all final 
state nucleons that are produced in the interaction $T=\sum_i T_i$ was measured 
and spectrum of NCE events $dN_{NCE}/dT$ was reconstructed as a function 
$T\approx\omega$. Assuming that the target nucleon is at rest, $Q^2_{QE}$ was 
determined for each event as
\begin{eqnarray}
\label{Eq.16}
Q^2_{QE}=2mT=2m\sum_i T_i.
\end{eqnarray}

MiniBooNE reported the NCE differential cross section in the range of 
$Q^2_{QE}$ from 0.1 to about 1.65 (GeV/c)$^2$. This differential cross section 
distribution was fitted with a dipole axial form factor and best fit for 
$M_A=1.39\pm0.11$~GeV was obtained. Using the data from charged-current 
neutrino interaction sample, the $NCE/CCQE$ cross sections ratio as a function 
of $Q^2_{QE}$ was measured. One should understand that, in fact, the NCE 
flux-averaged differential cross section $d\sigma/dT$ was measured, which 
under the assumption of a scattering on free nucleon, was recalculated as 
$d\sigma/dQ^2_{QE}=(d\sigma/dT)/2m$. The details of the unfolding procedure in 
measurement of the cross section can be found in Ref.~\cite{Denis}.

We calculated the NCE flux-averaged inclusive $\langle d\sigma_{NCE}/dT\rangle$
 cross section in the framework of RDWIA approach that was recalculated as 
$\langle d\sigma_{NCE}/dQ^2_{QE}\rangle = \langle d\sigma_{NCE}/dT\rangle/2m$. 
The details of the calculation $\langle d\sigma_{NCE}/dT\rangle$ cross section 
are described in Appendix A.  
To extract a value for the parameter $M_A$ we calculated this cross section 
with the Booster Neutrino Beamline flux~\cite{MiniBf} using the $Q^2_{QE}$ 
bins $\Delta Q^2=Q^2_{i+1}-Q^2_i$ similar to~\cite{MiniB}  
\begin{eqnarray}
\label{Eq.17}
\left(\frac{d\sigma}{dQ^2_{QE}}\right)_i=
\frac{1}{\Delta Q^2}\int_{Q^2_i}^{Q^2_{i+1}}
\left\langle\frac{d\sigma}{dQ^2_{QE}}(Q^2_{QE})\right\rangle dQ^2_{QE} 
\end{eqnarray}
and NCE to CCQE cross section ratio 
\begin{eqnarray}
\label{Eq.18}
R_i=NCE/CCQE=
\left(d\sigma_{NCE}/dQ^2_{QE}\right)_i/\left(d\sigma_{CC}/dQ^2\right)_i, 
\end{eqnarray}
The CCQE differential cross section 
$d\sigma_{CC}/dQ^2$ was calculated in the RDWIA approach in Ref.~\cite{BAV4}. 
The fit to the extracted flux-averaged $\langle d\sigma_{NCE}/dQ^2_{QE}\rangle$ 
yield the parameter $M_A=1.28 \pm 0.5$~GeV. Fig.\ref{Fig.7} shows the MiniBooNE measured 
flux-averaged differential 
$d\sigma_{NCE}/dQ^2_{QE}$ cross section and $R=NCE/CCQE$ ratio~\cite{MiniB} as a 
function of $Q^2_{QE}$ compared with the RDWIA calculations with the value of 
$M_A=1.28$~GeV. The result obtained with $M_A=1.37$~GeV, that was extracted in 
Ref.~\cite{BAV4} from the fit to measured in Ref.~\cite{MiniBCC} 
flux-integrated CCQE cross section $d\sigma_{CC}/dQ^2$, also is shown. 

There is an overall agreement within errors between the RDWIA predictions 
and the MiniBooNE data at high $Q^2_{QE}$. However one should note that at 
$Q^2_{QE} \ge 0.4$ (GeV/c)$^2$ the inclusive cross section, as well as, 
$R=NCE/CCQE$ ratio, calculated with the value of $M_A=1.28$~GeV has a better 
agreement with data. At $Q^2_{QE} < 0.25$ (GeV/c)$^2$ the calculation 
underestimate both cross section and the NCE/CCQE ratio by 17\% or less.

\subsection{MiniBooNE $\nu p \to \nu p/\nu N \to \nu N$ differential cross 
section cross-section ratio: comparison with data}

In addition to the $\nu N \to \nu N$ differential cross-section, 
MiniBooNE has published the $\nu p \to \nu p$ to $\nu N \to \nu N$ ratio at
$Q^2 > 0.7\mbox{ GeV}^2$ (above the Cherenkov threshold for protons in mineral 
oil)~\cite{MiniB}.
This result is interesting, because it should be sensitive to the strange 
quark contribution to the axial form factor. MiniBooNE has reported the 
measurement $\Delta s = 0.08\pm 0.30$, based on a Nuance prediction
~\cite{NUANCE}.

\begin{figure*}
  \begin{center}
    \includegraphics[height=8cm,width=16cm]{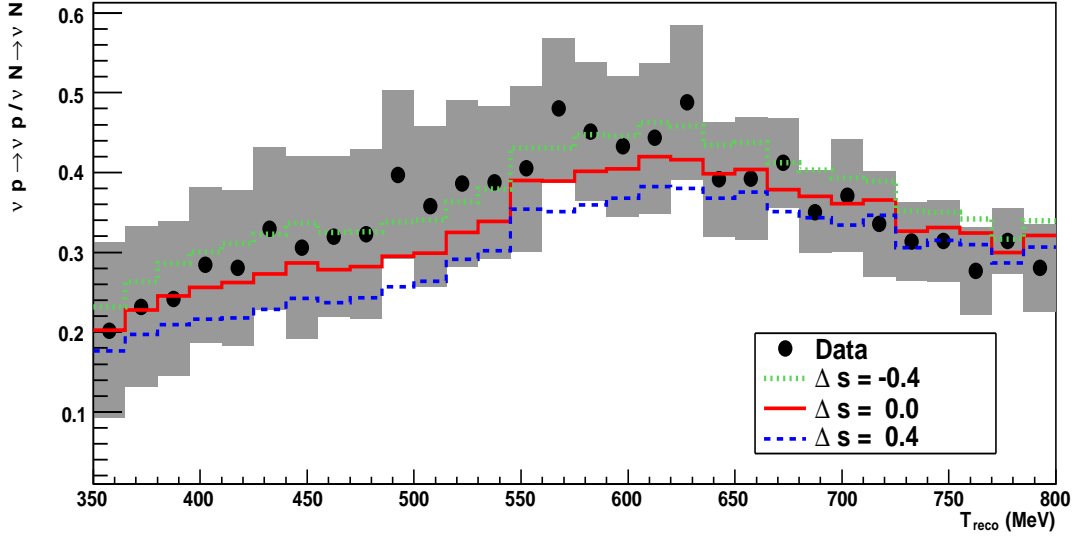}
  \end{center}
  \caption{\label{Fig.8}(Color online).
  MiniBooNE $\nu p \to \nu p/\nu N \to \nu N$ ratio as a function of $T_{rec}$.
  The prediction for $\Delta s=0.0$ and $\Delta s=0.4$ are shown as red solid 
  and blue dashed histograms, respectively. The MiniBooNE data are shown as 
  points with the total error bars on top of them.  
}
\end{figure*}

The $\nu p \to \nu p/\nu N \to \nu N$ ratio was reported as a function of the 
MiniBooNE reconstructed nucleon kinetic energy $T_{rec}$.
Also the migration matrices were published in Ref.\cite{MB_NCE_data_release}, 
which carry the detector resolution and efficiency information.
Using them one can smear the predicted cross-sections and obtain the predicted 
event rates as a function of $T_{rec}$. The procedure for carying out 
calculations of event rates in terms of the MiniBooNE reconstructed energy is 
described in an Appendix of Ref.\cite{Denis}. We performed the calculation of 
the $\nu p \to \nu p/\nu N \to \nu N$ ratio based on our neutrino interaction 
model and compared it to the MiniBooNE data. We have calculated our prediction 
of the event rates for different values of $\Delta s$ covering the range from 
$-0.4$ to 0.4. An example of the calculation is shown in Fig.\ref{Fig.8}.

Using the full error matrix for the ratio published in 
Ref.\cite{MB_NCE_data_release} we calculated the $\chi^2$ distribution between 
data and the MC. Our calculation leads to:
$$
\Delta s = -0.11\pm 0.36.
$$
with $\chi^2_{min}=33.4$ for 29 degrees of freedom. This result is consistent 
with all other measurements of $\Delta s$, including the one reported by 
MiniBooNE.
We show a calculated NCE cross-section and NCE/CCQE ratio with 
the values of $\Delta s=-0.11$ and $M_A=1.28$~GeV in Fig.\ref{Fig.7}.
As one can see the effect of strange quarks is small, but the agreement 
between data and our prediction does improve a little bit at 
low $Q^2$ region with $\Delta s=-0.11$.

\section{Conclusions}

In this article we study neutral-current elastic (anti)neutrino scattering on 
Carbon and $CH_2$ targets in the framework of RDWIA approach placing particular
 emphasis on nuclear effects. We calculated the NCE exclusive $d\sigma/dT$ 
cross sections for nucleon knockout in (anti)neutrino scattering on a Carbon 
target. The calculation presented in this paper are consistent with the RDWIA 
cross sections of Ref.~\cite{Martinez,Meucci1}.
 
We also calculated the $d\sigma/dQ^2$ inclusive cross sections for neutrino 
scattering on ${}^{12}$C, as well as, on free nucleon for different neutrino 
energies and estimated the range of $Q^2$ where nuclear effects on the shape of
 $Q^2$ distribution are negligible. We show that these effects in the CCQE and
 NCE scattering are similar. 

Using the RDWIA approach with the Booster Neutrino Beamline flux~\cite{MiniBf} 
we extracted axial mass from a ``shape-only'' fit of the measured 
flux-averaged $d\sigma/dQ^2_{QE}$ differential cross section. The extracted 
value of $M_A=1.28 \pm 0.05$ is in agreement within errors with the MiniBooNE 
result of $M_A=1.39 \pm 0.11$~GeV.    
There is a good overall agreement within errors in the range of 
0.25~(GeV/c)$^2<Q^2<1.65$~(GeV/c)$^2$ between RDWIA prediction and the 
MiniBooNE data: the measured MiniBooNE NCE flux-averaged differential cross 
section $d\sigma/dQ_{QE}^2$ on $CH_2$ and the $NCE/CCQE$ cross section ratio. 
However, in the range of low $Q^2 \le 0.25$ (GeV/c)$^2$ the calculations 
underestimate the measurements by 17\% or less.  

Using MiniBooNE data for the high energy 
$\nu p \to \nu p$ to $\nu N \to \nu N$ ratio the value of 
$\Delta s = -0.11\pm 0.36$ has been extracted based on our model,
which is consistent with other measurements of $\Delta s$.

We conclude that the RDWIA approach was successfully tested against neutrino 
CCQE and NCE scattering on ${}^{12}$C.   

\section*{Acknowledgments}

The authors greatly acknowledges to R.~Tayloe and G.P.~Zeller for fruitful 
discussions and a critical reading of the manuscript.   

\appendix

\section{Flux-averaged NCE inclusive cross section}
\label{A}

MiniBooNE measured the flux-averaged NCE differential cross section 
(per nucleon) on $CH_2$, averaged over three process: scattering off free 
protons in Hydrogen, bound protons in Carbon, and bound neutrons in Carbon. 
This cross section can be expressed as   
\begin{eqnarray}
\label{Eq.A1}
\frac{d\sigma_{\nu N}}{dQ^2_{QE}}=
\frac{1}{7}C_{\nu p,H}\frac{d\sigma_{p,H}}{dQ^2_{QE}}
+\frac{3}{7}C_{\nu p,C}\frac{d\sigma_{p,C}}{dQ^2_{QE}}
+\frac{3}{7}C_{\nu n,C}\frac{d\sigma_{n,C}}{dQ^2_{QE}},
\end{eqnarray}
where $d\sigma_{p,H}$ is the NCE cross section on free protons (per proton), 
$d\sigma_{p,C}$ is the cross section on bound protons (per proton), 
$d\sigma_{n,C}$ is the cross section on bound neutrons (per neutron), 
$C_{\nu p,H}, C_{\nu p,C}$, and $C_{\nu n, C}$ are the efficiency correction 
functions, that are given in Table IV of Ref.~\cite{MiniB}. 

In this paper the NCE inclusive cross sections for neutrino ($\nu_{\mu}+\nu_e$) 
scattering on bound proton $d\sigma_p/dQ^2_{QE}$ and on bound neutron 
$d\sigma_n/dQ^2_{QE}$ are calculated in the RDWIA approach. The flux-averaged 
$\langle d\sigma/d Q^2_{QE} \rangle$ cross section can be written as
\begin{eqnarray}
\label{Eq.A4}
\left\langle\frac{d\sigma}{dQ^2_{QE}}(Q^2_{QE})\right\rangle = \frac{1}{\Phi}
\int_{\varepsilon_{\min}}^{\varepsilon_{\max}}\frac{d\sigma_i}{dQ^2_{QE}}
(Q^2_{QE},\varepsilon_{\nu})[I_{\nu_{\mu}}(\varepsilon_{\nu})+
I_{\nu_e}(\varepsilon_{\nu})]d\varepsilon_{\nu},
\end{eqnarray}
where $I_{\nu_{\mu}} (I_{\nu_e})$ is the neutrino spectrum and $\Phi$ is the 
neutrino flux ($\nu_{\mu} + \nu_e$) in $\nu$- mode of beam, integrated 
over $0 \le \varepsilon_{\nu} \le 2.6$~GeV. This definition of the 
flux-averaged NCE inclusive cross section is similar to the definition in 
Ref.~\cite{MiniBCC} of  the flux-integrated CCQE differential cross section 
$d\sigma/dQ^2$. 


\end{document}